\documentclass[aps,prl,preprint,showpacs,showkeys,groupedaddress]{revtex4}
\newcommand{\W}{10cm}
\usepackage{graphicx}
\usepackage{amsmath}
\begin{document}
\title{Separation of suspended particles by arrays of obstacles in microfluidic devices}
\author{Zhigang Li}
\email{zli@jhu.edu}
\author{German Drazer}
\email{drazer@jhu.edu}
\affiliation{Department of Chemical and Biomolecular Engineering,
Johns Hopkins University, Baltimore, Maryland 21218, USA}
\date{\today}
\begin{abstract}
The stochastic transport of suspended particles through a periodic pattern of obstacles in
microfluidic devices is investigated by means of the Fokker-Planck equation.
Asymmetric arrays of obstacles have been shown to induce the continuous separation of DNA molecules
of different length.  The analysis presented here of the asymptotic distribution of particles
in a unit cell of these systems shows that separation is only possible in the presence of a driving 
force with a non-vanishing normal component at the surface of the solid obstacles. In addition, 
vector separation, in which different species move, in average, in different directions within the 
device, is driven
by differences on the force acting on the various particles and not by differences in the
diffusion coefficient. Monte-Carlo simulations performed for different particles and force
fields agree with the numerical solutions of the Fokker-Planck equation in the periodic system.
\end{abstract}
\maketitle

\begin{figure}
\includegraphics*[width=\W]{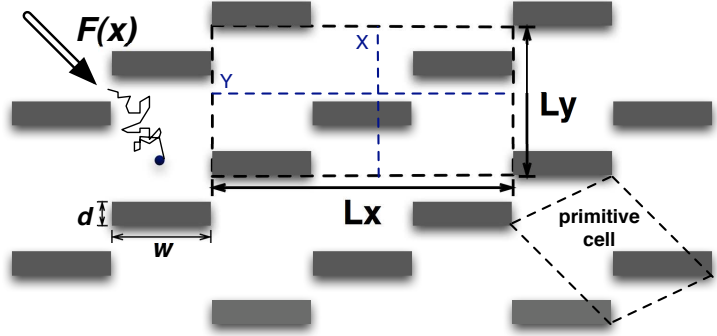}
\caption{\label{microdevice} Schematic view of a microfluidic device for the continuous-flow 
separation of suspended particles, equivalent to the one proposed in Ref.~\cite{DukeA98} and 
implemented experimentally in Refs.~\cite{ChouBTDCCCA99,HuangSTCSAC02}. 
The gray rectangles are solid obstacles. A suspended particle, represented
by a small sphere is transported through the sieving medium by an external field $\mathbf{F}(\mathbf{x})$. 
We show the unit cell used in this work, with dimensions $L_x$ and $L_y$ in $x$ and $y$ directions
respectively. We also show, on the right, the primitive cell of this 2D periodic system in dotted lines.}
\end{figure}

In recent years different separation strategies have been explored for the development of 
integrated microfluidic devices that could provide rapid and efficient fractionation of particles 
in solution. Particularly interesting are novel methods that take advantage of the intrinsic
Brownian motion of small particles suspended in solution, and the dependence of the magnitude of
the thermal motion on the size of the particles, to fractionate the initial suspension as
it is transported through the microfluidic device. Following the pioneering work 
by Austin and Volkmuth \cite{VolkmuthA92}, several experiments and theoretical investigations 
have shown the possibility to {\it rectify} the Brownian motion of suspended macromolecules 
(DNA in particular) by driving them through an asymmetric course of micrometer-scale obstacles 
\cite{Duke98,ChouBTDCCCA99,ChouABTCCCCDDHT00,LinHC05} 
(see Fig.~\ref{microdevice} for a schematic representation of the microdevice). 
These novel methods have the additional advantage that
separation occurs in a direction different (normal) to the average motion of the suspended particles,
thus allowing for continuous fractionation, in contrast to traditional separation
techniques for biological macromolecules such as electrophoresis \cite{GaalMV}.
However, the original explanation 
which assumes differences in the diffusivity of the suspended particles as the underlying mechanism leading to 
the observed separation
\cite{Duke98,Ertas98,ChouBTDCCCA99}
has been recently challenged by experimental observations and theoretical 
considerations \cite{HuangSTCSAC02,AustinDHSBD02}.
Here we investigate the transport of suspended particles in spatially periodic systems 
by means of the Fokker-Planck equation, as well as direct Monte-Carlo simulations,
and discuss the necessary conditions for separation to occur. 
In particular, we demonstrate that, in the absence of a driving force with a
normal component to the solid obstacles at the fluid-solid interface, 
the asymptotic concentration of particles is
uniform and no lateral separation is possible. We also show that
in the presence of a normal force at the surface of the obstacles, which induces a non-uniform 
concentration of particles, the {\it vector} separation of particles, 
in which different species migrate at different angles within the microdevice,
is driven by differences on the
force acting on the particles and not by differences on their mobility or diffusion coefficient.

The stochastic transport of suspended particles through an array of obstacles in the presence of 
an external force can be described by the Fokker-Planck equation for the probability 
density $P(\mathbf{x},t)$,
\begin{equation}
\label{fp}
\frac{\partial}{\partial t}P(\mathbf{x},t)
+\nabla \cdot \left\{\frac{\mathbf{F}(\mathbf{x})}{\eta} P(\mathbf{x},t)\right\}
-\frac{k_B T}{\eta} \nabla^2{P(\mathbf{x},t)}=0,
\end{equation}
where $\mathbf{x}=(x,y)$ is the position vector in the 2D microdevice, $k_B$ is the Boltzmann constant, 
$T$ is the temperature, $\mathbf{F}(\mathbf{x})$ is the force acting on the suspended particles, 
which we have assumed to be time-independent, and $\eta$ is the {\it friction coefficient} \cite{Risken}.
The diffusion coefficient is given by the well known Einstein relation $D=k_B T/\eta$, and for a hydrodynamic 
drag (friction) on a spherical particle of radius $a$ suspended in a viscous fluid of viscosity $\mu$ and moving at
low Reynolds numbers we arrive at the Stokes-Einstein relation $D=k_B T/6\pi\mu a$.

In general we are interested in computing the probability density flux describing the migration
of the suspended particles relative to the fixed sieve,
\begin{equation}
\label{flux}
\mathbf{J}(\mathbf{x},t)=
\frac{\mathbf{F}(\mathbf{x})}{\eta} P(\mathbf{x},t)
-\frac{k_B T}{\eta} \nabla{P(\mathbf{x},t)},
\end{equation}
and, in particular, to seek to determine whether the direction in which particles migrate is different for different
species, leading to {\it vector} separation \cite{DorfmanB01,DorfmanB02}.
In a periodic array of obstacles, such as the one shown in Fig.~\ref{microdevice}, 
it is convenient to introduce the {\it reduced} probability density and flux \cite{Reimann02},
\begin{eqnarray}
\label{reduced}
\tilde P(\mathbf{x},t)= \sum_{k_x=-\infty}^{+\infty}\sum_{k_y=-\infty}^{+\infty} P(x+k_xL_x,y+k_yL_y), \\
\mathbf{\tilde J}(\mathbf{x},t)= \sum_{k_x=-\infty}^{+\infty}\sum_{k_y=-\infty}^{+\infty}
\mathbf{J}(x+k_xL_x,y+k_yL_y),
\end{eqnarray}
where $\mathbf{k}=(k_x,k_y)$ identifies the specific cell within the periodic system and $L_x$, $L_y$ are the unit 
cell dimensions (see Fig.~\ref{microdevice}). The previous sums are expected to be convergent based on the 
normalization condition for $P(\mathbf{x},t)$.
Note that we have assumed an array extending to infinity in both $x$ and $y$ directions, or equivalently we are
neglecting finite size effects on the behavior of the transported species. 
Then, given the linearity of the Fokker-Planck equation, the convergence of the previous sums, and the
invariance under discrete translations in the periodic system, $(x\prime,y\prime)=(x+k_xL_x,y+k_yL_y)$, it is clear
that $\tilde P(\mathbf{x},t)$ is also a solution to the Fokker-Planck equation. In particular, $\tilde P$
satisfies Eq.~(\ref{fp}) with periodic boundary (and initial) conditions. 
It is clear that the suspended particles move between different cells but, due to the invariance of the system, it is 
as if a particle leaving a given cell was in fact returned at the equivalent point in the opposite face of the same cell, 
leading to the mentioned periodic boundary conditions in both 
$\tilde P(\mathbf{x},t)$ and  $\mathbf{\tilde J}(\mathbf{x},t)$. 
In addition, for transit times that are long enough so that the
system reaches its asymptotic state, we are interested in the long time behavior of 
the system, governed by the steady-state solution of the Fokker-Planck equation. 
Note that only the {\it reduced} probability reaches a steady state,
whereas the global probability density function has no meaningful asymptotic state \cite{Reimann02}.
Given an initial distribution of particles at the inlet of the microfluidic device, the steady state
is reached for transit times longer than the characteristic 
diffusive time required for a single particle to explore
the primitive unit cell. That is, when the transit time for a particle moving through $N$ cells, 
$\tau_C\approx N L/u$, where $u$ is the average velocity of the particle and $L$ is the dimension of the 
unit cell, is larger than the diffusive time required to explore the cell, 
$\tau_D\approx L^2/D$, where $D$ is the 
diffusion coefficient. The previous 
condition is satisfied for small {\it global} Peclet numbers, $\mathrm{Pe}=uL/ND\ll1$. 
In previous experimental works
the condition of steady state distribution, i.e. small Pe, is typically reached after the suspended 
particles are transported through a relatively small number of cells. In Ref.~\cite{HuangSTCSAC02}, for example,
the authors observe lateral separation for 48.5 kbp DNA molecules with $\mathrm{Pe}\approx 9/N$
($u\sim1\mu m$; $D\sim 0.64 \mu m^2/s$; $L\sim 6\mu m$) but no lateral separation for
411 bp DNA molecules with $\mathrm{Pe}\approx 0.5/N$ ($u\sim1\mu m/s$, $D\sim12\mu m^2/s$ and $L\sim6\mu m$).
In both cases, the steady state is in fact reached after the DNA molecules are transported 
through a few cells, compared to the total number of cells in the device, $N\sim1000$. 

In the presence of different species, such as suspended particles or DNA molecules of different size,
and assuming that the dilute approximation in which particle-particle interactions are neglected is
valid, we can describe the dynamics of the different particles by independently solving 
the steady state solution of the Fokker-Planck equation with periodic boundary conditions for each
of the species present in the system. In addition, for impermeable obstacles, the probability
density has to satisfy the zero-flux boundary condition at the surface of the 
solid obstacles. The governing equations then read,
\begin{eqnarray}
\label{steady}
\nabla \cdot \mathbf{\tilde J^\infty} = \nabla \cdot 
\left\{ \frac{\mathbf{F}(\mathbf{x})}{\eta} \tilde P^\infty(\mathbf{x}) 
-\frac{k_B T}{\eta} \nabla \tilde P^\infty(\mathbf{x}) 
\right\} =0\\
\label{j}
\left. \mathbf{\tilde J^\infty} \cdot \mathbf{n} \right |_{\partial s} = 0
\end{eqnarray}
where $\partial s$ is the surface of the obstacles and $\mathbf{n}$ is the local normal to the solid surface.
Once we obtain the steady state solutions, $\tilde P^{\infty}(\mathbf{x})$ 
and $\mathbf{\tilde J^{\infty}}(\mathbf{x})$, we can compute the average migration velocity 
(as well as its orientation), by computing the total probability flux across
the boundaries of the unit cell, that is,
\begin{equation}
\langle U_x \rangle = L_x \int_{0}^{L_y} dy ~ \tilde J^{\infty}_x,
\end{equation}
and the analogous equation for $\langle U_y \rangle$. 

\begin{figure}
\includegraphics*[width=\W]{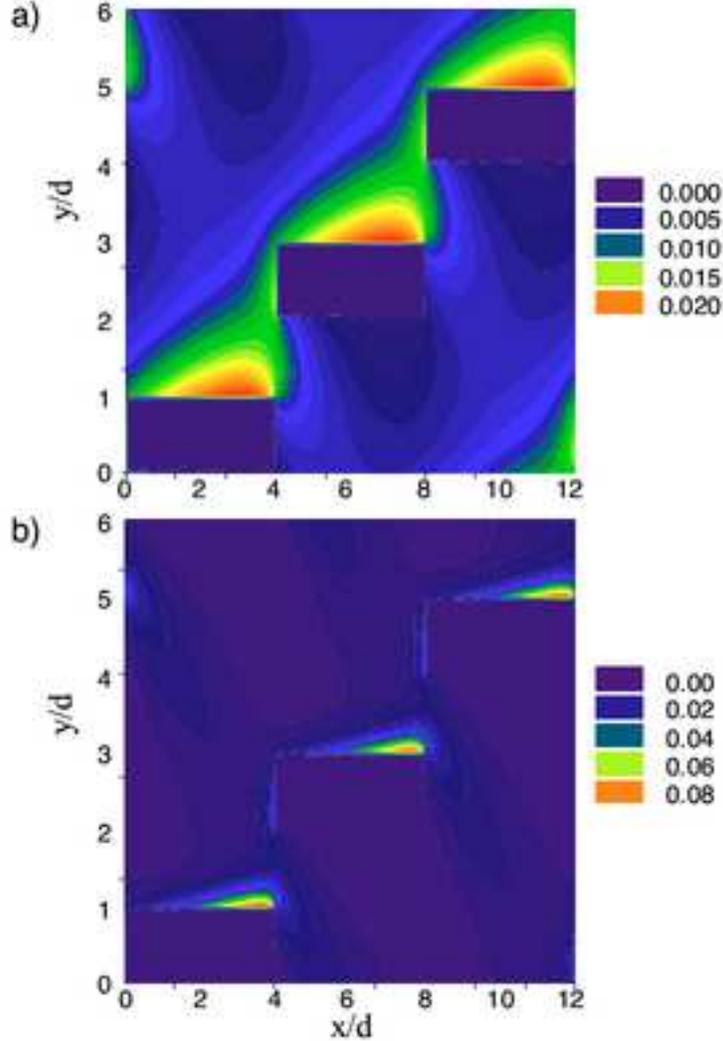}
\caption{\label{contour} 
Contour plots for the steady-state {\it reduced} probability density function
$\tilde P^{\infty}(\mathbf{x})$ inside a unit cell of the microfluidic device shown in Fig.~\ref{microdevice}. 
The force is spatially uniform and oriented at $45^\circ$, $F_x=F_y$. The difference between the
two plots is the magnitude of the {\em local} Peclet number, $\mathrm{Pe}=F\ell/k_BT$: 
a) $\mathrm{Pe}=1.12$ and b) $\mathrm{Pe}=5.58$.}
\end{figure}

Let us now consider two cases, classified according to the properties of the force field, that are relevant 
to the separation of particles in microfluidic devices.
{\em Case I: A solenoidal force field with no force normal to the solid obstacles.} This is probably the most common situation
in experiments. For example, if one uses electrophoretic driving forces with obstacles that
are impermeable to the ionic solution then the electric field outside the 
Debye double-layer (which is typically only a few nanometers thick) will be parallel to the solid surfaces as
the ions carried by the electric field move around the obstacles. This is also the case for hydrodynamic 
convection in pressure driven flows. In fact, in order to describe the motion of an advected particle 
we can replace $\mathbf{F}/\eta$ 
by the local fluid velocity $\mathbf{v}$, and again the normal component vanishes at the surface of the 
impermeable obstacles.
Unfortunately, the steady state in these cases is a uniform concentration of particles
in the unit cell, independent of the details of the forcing field or the physical properties
of the suspended species (Note that the force field satisfies $\nabla \cdot \mathbf{F}=0$). 
This is due to the fact that, since there is no force
towards the obstacles at the solid surface, there is no concentration gradient developing at the interfaces.
Therefore, the average velocity will be co-linear with the coarse-grained global force applied
on the system and no vector separation is possible,
\begin{equation}
\langle \mathbf{U} \rangle= 
\frac{1}{\eta} \left\{ \frac{1}{L_x L_y}\int_{0}^{L_x} \int_{0}^{L_y} dx dy \mathbf{F}(\mathbf{x}) \right\}.
\end{equation}
If the magnitude of the force divided by the mobility of the particles,
$F/\eta$, is different for different species, 
then the particles would move at different absolute velocities, but they will still move in the same direction 
(relative orientation within the device).
{\em Case II: Force penetration into the solid obstacles.} 
In Fig.~\ref{contour} we show the steady-state solution of the probability density 
for a spatially uniform force oriented at $45^\circ$ ($F_x=F_y$).
It is clear that concentration gradients develop close to the solid obstacles, which increase
with the magnitude of the external force $F$. 
It is possible, in fact, to render Eqs.~(\ref{steady}) and (\ref{j}) nondimensional
using the {\em local} Peclet number, ${\mathrm Pe}=(U \ell)/D={F \ell/ (k_B T)}$, where $\ell$ is a
characteristic length scale for the obstacles. 
Figure~\ref{contour} presents results for two different Peclet numbers, $\mathrm{Pe}\sim1$ and
$\mathrm{Pe}\sim5$, where we used $\ell=d$ (see Fig.~\ref{microdevice}). 
These Peclet numbers are representative of some of the recent experiments on the transport of
DNA molecules in microdevices. For example, 
in Ref.~\cite{HuangSTCSAC02} the authors report good separation for local Peclet numbers
in the range $2.1<\mathrm{Pe}<5.1$ and in Ref.~\cite{ChouBTDCCCA99} the authors observe 
the best separation performance for $\mathrm{Pe}\sim3.5$. 
On the other hand, the asymptotic solution in the limit of high Peclet numbers is singular, 
with high concentration gradients present in a thin boundary layer of thickness $\xi\sim{(k_B T)/F_x}$.
In this limit, the assumption of point-particles will no longer be valid for $\xi \lesssim a$.

\begin{figure}
\includegraphics*[width=\W]{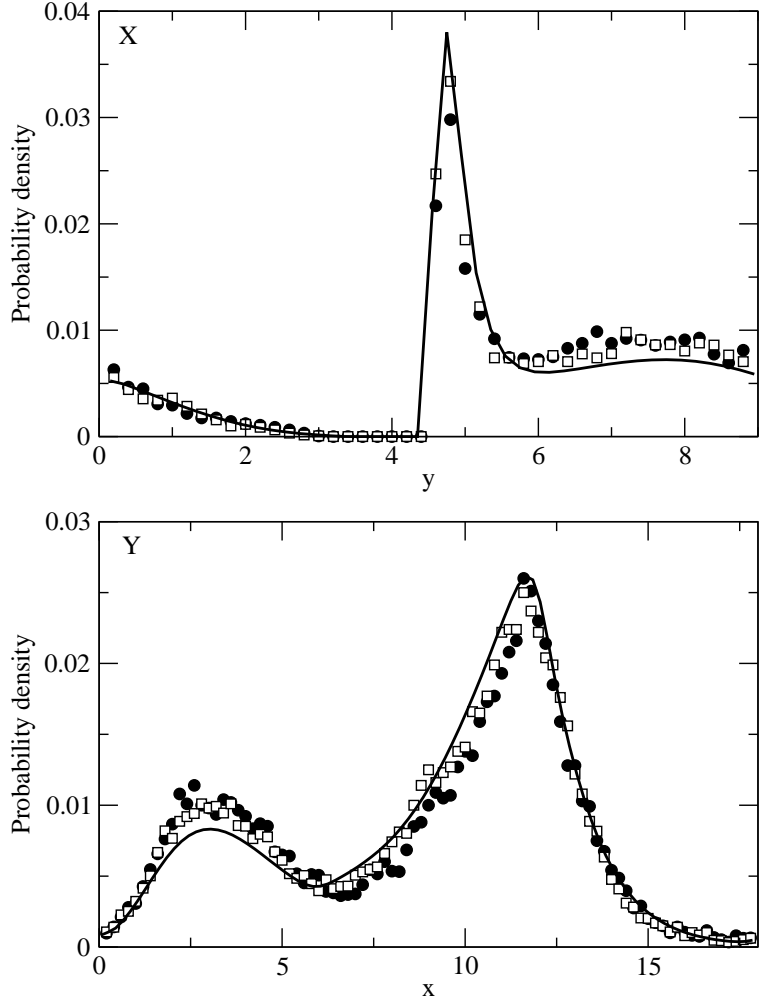}
\caption{\label{cmp} Probability distribution function at the two perpendicular lines, 
$X$ and $Y$, in the unit cell (see Fig.~\ref{microdevice}). The force is spatially uniform and oriented at
$45^{\circ}$, and the Peclet number is $\mathrm{Pe}=5.58$.
The solid line corresponds to the solution to Eqs.~(\ref{steady}) and (\ref{j}) presented in
Fig.~\ref{contour}b. 
The open and solid points correspond to Monte-Carlo simulations for two types of particles, which
only differ on their mobility, and thus their diffusion coefficient, by a factor of $10$.}
\end{figure}

The migration velocity will in general depend on the specific force field acting on the particles,
but we can, however, obtain some interesting results by considering the properties of the
steady state solution to Eqs.~(\ref{steady}) and (\ref{j}). First of all, for a given force field
$\mathbf{F}(\mathbf{x})$, the steady-state probability density function, $\tilde P^{\infty}$,
is independent of the mobility of the particles and, therefore, independent of their
diffusivity.  As a consequence, if we have 
species $A$ and $B$ with mobilities $\eta_A^{-1}$ and $\eta_B^{-1}$ their average migration velocities 
will be proportional to each other, 
$\langle \mathbf{U}_A \rangle = (\eta_B/\eta_A) \langle \mathbf{U}_B \rangle$, and there will be no
vector separation in this case. 
In order to test this result, we compare the solution to Eqs~(\ref{steady}) and (\ref{j}) with Monte-Carlo simulations
for the motion of suspended particles. We simulated the motion of two species of particles which 
experience the same force field but have different diffusivities, e.g. the case of equally charged
particles of different size driven by an external electric field. First, we simulated the motion
under an electric field that does not penetrate the obstacles and obtained, as expected, a uniform
distribution of particles. Then, we simulated the motion under the action of a spatially uniform force 
and, in Fig.~\ref{cmp}, we compare the steady-state solution presented in Fig.~\ref{contour} with the 
simulation results. It is clear 
that the simulations results agree well with the steady state solution to the Fokker-Planck equations
with periodic boundary conditions. We can also see that the asymptotic distribution of the two different
types of particles is the same, independent of the value of their mobility. Then, particles with
an order of magnitude different diffusivities will migrate in the same average direction.
Fortunately, in the case of DNA molecules, although the
electrophoretic velocity is independent of size, the electric force acting on DNA molecules of 
different length is not the same ($\lambda \approx 0.3 e^{-}/\mbox{\AA}$
is the effective charge per length
of the DNA molecule \cite{VolkmuthA92,VolkmuthDWAS94,Ertas98}). Therefore, it could be possible to achieve
vector separation by creating an electric field with a non-vanishing normal component at the solid surfaces
using, for example, permeable obstacles.
In the previous analysis, however, we neglected the hydrodynamic interactions between the suspended
particles and the solid obstacles, which would lead to a position-dependent mobility \cite{KimK}.
Then, for large enough particles compared to the characteristic size of the obstacles, there might be
separation induced by the difference in permeability of particles of different size \cite{DorfmanB01,DorfmanB02}.

We showed that the stochastic transport of particles through periodic patterns of obstacles
can be described by the asymptotic steady-state solution of the Fokker-Planck equation applied to a unit cell
with periodic boundary conditions and appropriate condition at the surface of the solid obstacles.
We demonstrated that, in order to obtain {\it vector} separation of particles, a non-vanishing driving force
normal to the obstacles is required at the fluid-solid interface. We also showed that the underlying cause 
driving separation is the difference
in the force experienced by different species and not the differences in the diffusion coefficient. 
These
results could provide guidance for future developments in the design of microfluidic devices, such as
the use of permeable obstacles in combination with electrophoretic forcing of the particles. In addition,
the numerical solution of the Fokker-Planck equation in different geometries could then be used to
optimize the design of such microdevices. It is interesting to note that an analogous formalism to the one 
used here has been developed to study transport in periodic porous
media ({\em discontinuous systems}) and a nice and thorough discussion can be found in chapter 4 of
Ref.~\cite{BrennerE}. There, the authors show that the use of such method goes beyond the study of particle migration and
also allows the direct computation of the dispersion coefficient, responsible for the broadening of
an injected band of suspended particles, an important parameter affecting the resolution in separation
devices. Finally, we mention that we are currently considering the effect of transient dynamics before 
steady state is reached, which could lead to permanent separation in short devices.

\bibliography{articles,mybooks,book,mios}
\bibliographystyle{apsrev}

\end{document}